\documentclass[10pt,letterpaper]{article}
\usepackage[top=0.85in,left=1.0in,footskip=0.75in]{geometry}

\usepackage{changepage}

\usepackage[utf8x]{inputenc}

\usepackage{textcomp,marvosym}

\usepackage{amsmath,amssymb}

\usepackage{cite}

\usepackage{nameref,hyperref}

\usepackage[right]{lineno}

\usepackage{microtype}
\DisableLigatures[f]{encoding = *, family = * }

\raggedright
\usepackage[aboveskip=1pt,labelfont=bf,labelsep=period,justification=raggedright,singlelinecheck=off]{caption}

\makeatletter
\renewcommand{\@biblabel}[1]{\quad#1.}
\makeatother

\date{}

\usepackage{lastpage}
\usepackage{fancyhdr}
\usepackage{graphicx}
\usepackage{epstopdf}
\usepackage{color}
\usepackage[usenames]{xcolor}
\usepackage{soul}
\usepackage{mathtools}
\usepackage{sidecap}
\usepackage{array}
\usepackage{longtable}
\usepackage{pdflscape} 

\usepackage{floatrow}
\DeclareFloatFont{small}{\fontsize{9}{11}\selectfont}
\definecolor{Rred}{RGB}{178, 24, 43}
\definecolor{Rorange}{RGB}{214, 96, 77}
\definecolor{Rlrdorange}{RGB}{244, 165, 130}
\definecolor{Rltorange}{RGB}{253, 219, 199}
\definecolor{Reggshell}{RGB}{247, 247, 247}
\definecolor{RCHIsky}{RGB}{209, 229, 240}
\definecolor{RCHIltblue}{RGB}{146, 197, 222}
\definecolor{RCHImedblue}{RGB}{67, 147, 195}
\definecolor{RCHIblue}{RGB}{33,102, 172}

\begin{document}

{\LARGE
\begin{center}
\textbf{Properties of Healthcare Teaming Networks as a Function of Network Construction Algorithms}
\end{center}}

\begin{flushleft}
\vspace{1mm}

\begin{center}
Martin S. Zand\textsuperscript{1,2,3*},
Melissa Trayhan\textsuperscript{1,3},
Samir A. Farooq\textsuperscript{1,3},
Christopher Fucile\textsuperscript{1,4},

Gourab Ghoshal\textsuperscript{5},
Robert White\textsuperscript{1,3}, 
Caroline M. Quill\textsuperscript{1,6},
Alexander Rosenberg\textsuperscript{1,4},

Hugo Serrano\textsuperscript{5},
Hassan Chafi\textsuperscript{7},
Timothy Boudreau \textsuperscript{7}
\end{center}

\begin{list}{}{\setlength\itemindent{-\leftmargin} \itemsep0em}

\item \textsuperscript{\textbf{1}} Rochester Center for Health Informatics, University of Rochester Medical Center, Rochester, NY, USA             
\item \textsuperscript{\textbf{2}} Clinical Translational Science Institute, University of Rochester Medical Center, Rochester, NY, USA
\item \textsuperscript{\textbf{3}} Department of Medicine, Division of Nephrology, University of Rochester Medical Center, Rochester, NY, USA
\item \textsuperscript{\textbf{4}} Department of Medicine, Division of Allergy, Immunology and Rheumatology, University of Rochester Medical Center, Rochester, NY, USA
\item \textsuperscript{\textbf{5}} Department of Physics, University of Rochester, Rochester, NY, USA 
\item \textsuperscript{\textbf{6}} Department of Medicine, Division of Pulmonary and Critical Care Medicine, University of Rochester Medical Center, Rochester, NY, USA
\item \textsuperscript{\textbf{7}} Oracle Labs, Belmont, CA, USA
\end{list}
\bigskip

* martin\textunderscore zand@urmc.rochester.edu

\end{flushleft}
\section*{Abstract}
Network models of healthcare systems can be used to examine how providers collaborate, communicate, refer patients to each other, and to map how patients traverse the network of providers. Most healthcare service network models have been constructed from patient claims data, using billing claims to link a patient with a specific provider in time.  The data sets can be quite large (\(10^6$--$10^8$ individual claims per year), making standard methods for network construction computationally challenging and thus requiring the use of alternate construction algorithms. While these alternate methods have seen increasing use in generating healthcare networks, there is little to no literature comparing the differences in the structural properties of the generated networks, which as we demonstrate, can be dramatically different.  To address this issue, we compared the properties of healthcare networks constructed using different algorithms from 2013 Medicare Part B outpatient claims data. Three different algorithms were compared: binning, sliding frame, and trace-route. Unipartite networks linking either providers or healthcare organizations by shared patients were built using each method. We find that each algorithm produced networks with substantially different topological properties, as reflected by numbers of edges, network density, assortativity, clustering coefficients and other structural measures. Provider networks adhered to a power law, while organization networks were best fit by a power law with exponential cutoff.  Censoring networks to exclude edges with less than 11 shared patients, a common de-identification practice for healthcare network data, markedly reduced edge numbers and network density, and greatly altered measures of vertex prominence such as the betweenness centrality. Data analysis identified patterns in the distance patients travel between network providers, and a striking set of teaming relationships between providers in the Northeast United States and Florida, likely due to seasonal residence patterns of Medicare beneficiaries.  We conclude that the choice of network construction algorithm is critical for healthcare network analysis, and discuss the implications of our findings for selecting the algorithm best suited to the type of analysis to be performed. 

\section*{Introduction}

Network science can provide key insights into healthcare systems including patient referral patterns\cite{RN48, RN40, RN47, RN57, RN46, RN18, RN44, RN32, RN51, RN39}, provider communities associated with better healthcare outcomes, or specific drug prescribing patterns\cite{RN20, RN42, RN16}.  Network analysis is particularly useful for studying healthcare delivery by organizations (e.g. private practice groups and hospital networks) and providers (physicians, nurse practitioners, physical therapists, etc.). The research questions suited to network science methods typically fall into three categories:  1) network topology; 2) patient flow; and 3) provider clustering.   Network topology questions include investigations of network structure and properties, such as the effect of the rules and constraints under which provider teams organize (i.e. referral bias, geographic proximity, insurance network restrictions)\cite{RN15} or identifying providers with high levels of influence.  In contrast, questions about network flow address patterns of patient movement, network capacity and dynamic instability (e.g. how influenza epidemics or hospital closures affect network capacity). Provider clustering can identify highly collaborative groups of providers associated with specific patient outcomes.  Such work is crucial for identifying provider groups (e.g. communities, k-cliques or k-clans) with good outcomes for patients with complex conditions, such as cancer, heart failure or kidney disease\cite{RN33, RN26, RN29}.

All of these inquiries start by building a healthcare network model, with vertices representing providers or healthcare organizations, linked by edges representing the strength of the connection, generally the number of shared patients\cite{RN48, RN58, RN32}.  Several types of network construction algorithms exist, each with specific applications.  For example, matrix algebra methods are often used to construct social networks from moderate sized data sets. In contrast, trace-route mapping algorithms are used to create network representations for the study of network flow (e.g. digital information, transportation, supply chains). These types of methods have been used to map the flow of information across the internet \cite{RN113, RN114, RN116}, through social networks\cite{RN133, RN137, RN146}, and metabolite flow in bacterial biochemical pathways\cite{RN148}.  However, studies of which algorithms are best suited to constructing healthcare networks are lacking in the literature.

The most basic algorithmic method of healthcare network construction is to find all the instances where a specific provider $x_i$ sees a patient $y_i$ at least once, create a large patient-by-provider table, and then transform it into a provider-provider network (PPN) with each vertex representing a provider and each weighted edge representing the number of shared patients between the two providers. This network construction method uses no temporal information about the direction of the provider-patient visits, but merely specifies the volume of shared patients over the sampling period. The resulting networks are well suited to identify provider teams or links between healthcare organizations, organization-organization networks (OON), that share large numbers of patients.  In contrast, study of patient flow between providers requires building a network representation that captures sequence of patient visits to providers using algorithms that build networks based on the temporal ordering of provider visits. For example, adding up all of the visits where a patient goes from provider $P_i \rightarrow P_j$, with the time of the visits such that $t_i \le t_j$ and doing this for all providers in a data set, yields such a flow network that describes how patients move through the healthcare provider network, and how they are linked.  One such method (the sliding frame algorithm is described in detail in the \textit{Methods section}, has been used by the United States Center for Medicare Services to construct the annual United States Medicare Physician Referral Datasets \cite{RN118,RN980,RN67, RN71}, and to determine referral volumes from general practitioners to specialists\cite{RN15, RN55, RN23}.  

Crucial to healthcare network analysis is selecting a network construction algorithm appropriate for the analytic goal, and understanding how the algorithm affects the results obtained from analysis. Despite the increasing use of network models to improve healthcare delivery and outcomes\cite{RN48,RN57,RN34,RN45,RN46,RN50}, rigorous published reports comparing the networks constructed with different algorithms are lacking. There is also little guidance addressing the choice algorithms for different types of analyses.   Different methods are likely to result in networks with different elements (e.g. numbers of vertices and edges), topology, and properties (e.g. vertex degree and centrality distributions, edge weights, communities identified). In addition, the relationship between network topology and meaning is complex and tightly linked.  For example, do edges represent referrals, the act of a sending a patient to a provider for a specific consultation?  What algorithms create networks best suited to identify teaming, the grouping of providers that share many common patients and collaborate on their care?  Thus, the choice of network construction algorithm may have significant implications for network properties and inferred meaning of network topology.

The choice of network construction algorithm is also affected by the size of the data set and the computational complexity and memory required for the calculations \cite{RN132,RN133}.  Healthcare networks are generally constructed from data with a simple data structure, each record containing the date, the provider’s unique identifier, type of event (e.g. visit, admission, lab test), and the organization of which the provider is a member (e.g. practice group, healthcare system).  This data can then be linked to provider and patient demographic features and outcomes.  The Medicare Part B annual data sets contain $\sim$150--200 million individual claims from 800,000 providers for $\sim$25--40 million patients, giving $\sim 2.0 \times 10^{13}$ data elements.  This makes in-memory storage difficult, and network construction by conventional matrix dot product calculations computationally expensive \cite{RN109,RN110}. Algorithmic approaches, however, can provide an efficient and parallelizable implementation of network construction. 

Motivated by these issues, we characterize the consequences of choosing particular network-generating algorithms on the study of healthcare delivery networks. In the following manuscript, we compare the network topology and properties of Medicare PPN and OON constructed from the same primary data set using three different algorithms, and discuss the implications of each method for healthcare network analysis. 

\section*{Materials and Methods}
\subsection*{Human Subjects Protection}
Research data were coded such that patients could not be identified directly, in compliance with the Department of Health and Human Services Regulations for the Protection of Human Subjects (45 CFR 46.101(b)(4)). The analysis presented here is  compliant with Center for Medicare Services (CMS) current cell size suppression policy as well as all data exclusivity requirements contained in the CMS Limited Data Set Data Use Agreement.  This project was approved by the University of Rochester Institutional Review Board under the ``exempt" category.

\subsection*{Data Sources}
Network construction  algorithms were initially developed in PERL 5.22.1 using the CMS 2008-2010 Data Entrepreneurs’ Center for Medicare Services Outpatient Claims DE-SynPUF (DE-SynPUF)\cite{RN120}. This file contains institutional outpatient annual claim information for a 5\% sample of Medicare members’ outpatient Part B claims (i.e. 5\% of all claims randomly sampled) spanning from 2008 to 2010.  Each of the 15.8 million records in the DE-SynPUF file is a synthetic outpatient claim.   The DE-SynPUF dataset is publicly available for developers to test algorithms\cite{RN120}.  After development, the algorithms were tested and validated on the 2013 Medicare Outpatient Claims Limited Data Set (LDS) obtained from the Center for Medicare Services Research Assistance Data Center (ResDAC)\cite{RN121}. These combined files contain over 160 million Medicare fee-for-service claims data submitted by all organization and individual outpatient healthcare service providers between January 1, 2013 through December 31, 2014, along with a unique claim identifier number, dates of service, and unique National Provider Identifier numbers (NPIs). 

Provider information was abstracted from the National Plan and Provider Enumeration System (NPPES) data file\cite{RN122}  This file contains identifier information for all current and past United States licensed healthcare provider and organizations, each linked to a unique NPI number, and associated provider locations, demographics, and medical specialty information. We used the version from August, 2015, containing 4,763,891 NPI numbers of both organizations and individual providers.  All NPI numbers were checked for validity using the Luhn algorithm\cite{RN72}. Provider locations matched to a geo-coded NPPES downloadable file from July, 2014 by the North American Association of Central Cancer Registries (NAACCR) \cite{RN130}.  The file contains 4,180,737 NPI numbers and associated address, of which only 309 are lacking enough data to accurately geocode, and 179,614 are geocoded only at the zip code centroid level.  Geocoding is to the second decimal point, giving a spatial resolution of 1.1 km (~0.88 miles).

\subsubsection*{Data and Algorithm Availability}

The Center for Medicare Services Outpatient Claims DE-SynPUF (DE-SynPUF)\cite{RN120} test set is publicly available from the CMS web site.  The full 2013 Medicare Part B Limited Data Set for Medicare claims can be obtained from the Center for Medicare Services. This data is bound by a privacy and limited distribution agreement, as well as HIPAA regulations, and thus cannot be made public with this manuscript.  However, the files can be requested from the Center for Medicare Services by individual investigators and used to reproduce our findings.  Release of the derived networks is also limited by Medicare requirements to remove nodes and edges where the total number of shared patients $\leq11$.  This restriction is in place to prevent identification of individual patients based on a small number of visits to a unique combination of geographically identifiable providers \cite{RN118}.  Censored networks are available from figshare.com (10.6084/m9.figshare.3833943).  Network construction algorithms can be downloaded from figshare.com (doi 10.6084/m9.figshare.3837717), and are released under a GPL 3.0 license. 

\subsection*{Network Construction Algorithms}
\begin{figure}[!ht]
\centering
\includegraphics[width=.6\linewidth]{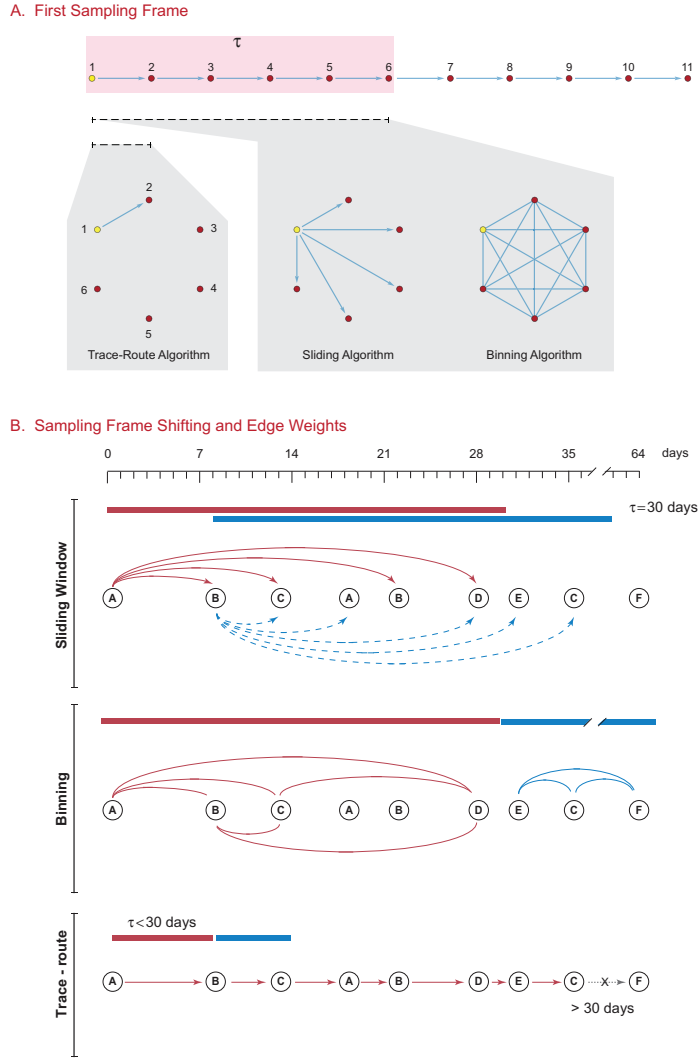}
\caption{{\bf Edge construction algorithms for healthcare teaming networks.}  Each vertex represents a provider, with the index provider vertex in yellow.  The collection of providers is for a single patient.  (A) The brackets show how the time frame $\tau=$30 days is applied to a series of temporally ordered provider visits.  The corresponding graphs show the edges that would be constructed between the provider vertices by each algorithm for that first iteration:  a single directed edge for the trace-route algorithm, a set of directed edges (e.g. a star graph) for the sliding algorithm, and a complete graph with each vertex connected to all other vertices for the binning method.  This process is repeated for each patient by shifting the sampling frame through the ordered visits for each patient. (B) Sampling frame shifting and edge weight construction.  How each algorithm shifts the sampling frame $\tau$ through the series of provider visits is shown here, and the degree of shift for the next interval is shown by the new brackets.  Please refer to the text for a discussion of edge weight calculations.}
\label{fig1}
\end{figure}

We constructed both provider and organization teaming graphs using three different algorithms, which we refer to as: (1) binning; (2) sliding frame; and (3) trace-route methods, adapting the terminology from Karimi and Holme (2013), who describes such frames in the context of dynamic networks \cite{RN131}. The essential features of the algorithms are illustrated in Fig~\ref{fig1}, with the subsequent mathematical description below and associated nomenclature listed in Table~\ref{table1}. 

\hspace{1mm}
\begin{table}[!ht]
\setlength{\tabcolsep}{2pt}
\caption{\vspace{1mm} \bf{Nomenclature}}
\begin{tabular}{l@{\hskip 1cm} l}
\hline
\\ 
\vspace{2mm}Symbol & Definition \\
\hline  \\
\(v_i$ & Vertex (organization or provider) where $i$ refers to identity of vertex type.\\
\(k_i$ & Degree of vertex $i$\\
\(e_{v_k \rightarrow v_l} ^{j} $ & Directed edge  between vertex $v_k$ and $v_l$ for patient $j$ ($\leftrightarrow$ when undirected).\\ 
$E_{v_k\rightarrow v_l}$ & Edge between $v_k$ and $v_l$ over all patients.\\
\(\omega_{v_k \rightarrow v_l}^{j}$ & Edge weight of $e_{v_k\rightarrow v_l}^{j}\)\\ 
\(\Omega_{v_k \rightarrow v_l}$ & Edge weight of $\omega_{v_k \rightarrow v_l} ^j$ over all patients\\ 
\bf{$V$, $E$, $\Omega$, $\textbf{P}$ ,\(\textbf{C}$}  & Respectively, sets of all vertices, edges, edge weights, patients, and claims\\
\(t_i$ & Temporal instance of vertex $i$\\
\hline
\end{tabular}
\label{table1}
\end{table}


\subsubsection*{Common algorithm features}
All three algorithms selected data with a temporal visit proximity frame, only counting graph edges if the provider visits were within $t_{\text{visit}} \leq \tau$ days of each other, where $\tau$ is the frame interval.  For each method described below, let $\mathbf{C}$ be the collection of all Medicare claims such that $c_{i}(p,v,t)$ where $p$ is the patient, $v$ is the provider, and $t_i$ the time of the patient visit for $i=\{1,2,3,..s\}$ where $s$ is the number of claims.  For all three algorithms, we consider all the claims for each patient $p_j$ in the claims set $\mathbf{C}$, where $j=\{1,2,3,..a\}$ and $a$ is the number of individual patients.  Each claim records the time of the patient visit to one provider.  Claims are grouped by patient, and then sorted in ascending temporal order.  The subsequent steps differ by algorithm, and are described below.  

\subsubsection*{Binning network construction algorithm}
The binning algorithm creates non-directed provider-provider network graphs.  It is essentially an algorithmic implementation of the unipartite projection of a bipartitie adjacency matrix, with the potential advantage (for very large claims data sets) of not requiring in-memory matrix dot products for network creation. For the binning method, we start with the subsets of claims for each individual patient $p_j$, where $j=\{1,2,3,..a\}$ and $a$ is the number of individual patients. Iterating over each patient, we consider all the claims in each set and create edges such that:   

\begin{eqnarray}
\label{eq:edgeCr2}
e_{v_k \leftrightarrow v_l} ^j\; \text{   if } \;
\begin{dcases}
   | t_l - t_k| \leq \tau \\
    v_k \neq v_l  \\
\end{dcases}
\end{eqnarray}
where $t_i$ refers to the time ordered instance of a particular claim ($i$ being the identity of the provider).
The total number of edges between vertices is
\begin{eqnarray}
\label{eq:edgeCr3}
E_{v_k \leftrightarrow v_l} = \sum_{j = 1}^{a} e_{v_k \leftrightarrow v_l} ^j
\end{eqnarray}
where $t_i$ refers to the time ordered instance of a particular claim ($i$ being the identity of the provider).
\noindent For individual patient edge weights within $\tau$, only the first interaction for any given provider-provider pair and patient $p_j$ is counted. For example, if $e_{v_k \rightarrow v_l}^j$ occurs 4 times within the frame $\tau$, the weight is only counted once, or succinctly
\begin{equation}
 \omega_{v_k \rightarrow v_l}^{j} = 1 \quad \text{if} \quad e_{v_k \leftrightarrow v_l} ^j \neq 0.
 \end{equation}
The final edge weights $\bf{\Omega}$ are calculated in a similar way as in the total edges by summing over all patients thus
\begin{eqnarray}
\label{eq:edgeWtA}
\Omega_{v_k \leftrightarrow v_l} = \sum_{j=1}^{a} \omega_{{v_k \leftrightarrow v_l}}^j
\end{eqnarray}

\subsubsection*{Sliding frame network construction algorithm}

Time directed network construction algorithms are designed to capture information contained in the temporal relationship of provider visits and used to build directed unipartite graphs.  We first describe the \textit{sliding frame algorithm}, one algorithm of this class.  The sliding frame algorithm is similar to the current algorithm used to create the publicly available Medicare physician shared patient data sets available from the Center for Medicare Services website \cite{RN123, RN980}.

In this setting, two providers are connected with a \emph{directed} edge if two claims for visits with the same patient occur within time $\tau$ when claims are sorted by increasing order of time. That is:
\begin{eqnarray}
\label{eq:edgeConst}
e_{v_k \rightarrow v_l}\; \text{   if } \;
\begin{dcases}
    0 < t_l - t_k \leq \tau \\
    v_k \neq v_l
\end{dcases}
\end{eqnarray}
Edge weights can either be are assigned as in the binning method described above, or edges can be be weighted by incidence within the frame such that $\omega^{p_j}_{v_k \rightarrow v_l}$ is the number of occurrences of $e_{v_k \rightarrow v_l}$ for patient $p_j$ within all frames $\tau$. The final edge weight $\Omega_{v_k \rightarrow v_l}$ within the entire graph of all patients $\bf{P}$ is calculated by

\begin{eqnarray}
\label{eq:edgeWtB}
\Omega_{v_k \rightarrow v_l} = \sum_{j=1}^{a} \omega^{p_j}_{{v_k \rightarrow v_l}} 
\end{eqnarray}

\noindent where $a$ is the number of unique patients in $\bf{P}$. If edges are counted more than once for a patient, $\Omega$ represents the edge weight of shared patients.  In contrast, if edges are counted each time they occur between $\omega^{p_j}_{v_k \rightarrow v_l}$, they represent the total number of visits between providers of shared patients.

We refer to this method as the \textit{sliding frame algorithm} due to the sequential scanning for relationships within the frame period $\tau$. Others have proposed that this captures the urgency of patient referrals between providers, for example when $\tau = 30$, where patients are directed by one provider to receive care from a second provider for an urgent medical issue\cite{RN980}. The requirement for $ v_k \neq v_l$ excludes self-looping edges (e.g. sequential visits to the same provider). This has been proposed by some so that the network include only referrals \cite{RN980}.  

Once the edges are created for each for patient, edge instances are tallied to obtain the overall edge weights for the entire network.  Claims-weighted edges have the value of the total number of claims for patients shared by two providers summed over all shared patients.  In contrast, patient-weighted edges are the sum of the number shared patients between two providers irrespective of the number of claims.  The resulting provider-provider network graphs are weighted and directed. 

\subsubsection*{Trace-route network construction algorithm}
The trace-route algorithm is similar to that used to create a map of the internet, and traces the route of a patient through temporally sequential provider visits.    The edges reflect provider-provider connections by sequential patient visits, and the edge weights are the rates of patient flow from provider to provider through the network for the period $\tau$.   The edge creation conditions can be specified by:
\begin{eqnarray}
\label{eq:edgeCr1}
e_{v_k \rightarrow v_l}\; \text{   if } \;
\begin{dcases}
    0 < t_l - t_k \leq \tau \\
    c_{k} \; \text{and} \; c_{l} \; \text{are strictly sequential claims}
\end{dcases}
\end{eqnarray}

\noindent In contrast to other methods, self-loops are permitted such that $e_{v_k \rightarrow v_k} $ can be counted as an edge.  Self-loop structures are common in strict temporally sequential claims data and reflect the case where an individual returns for successive visits to the same provider to address an ongoing condition or follow up after a procedure. Calculation of edge weights is the same as the sliding frame algorithm as noted above in equation (4).

\subsection*{Network Comparison}
Network comparisons were performed using standard network metrics in \textit{Mathematica 10.4.1} or in Oracle PGX (see below).  The definition of most metrics can be found in the excellent review by Newman \cite{RN143}.  Network metrics used in this manuscript included:

\begin{itemize}
\item \textit{Component enumeration}: We enumerated the total number of vertices and edges within each network, and the largest connected component $(lco)$ \cite{RN143}.  These correspond to the total number of unique providers or organizations, and the connections via shared patients between them.

\item \textit{Network diameter $(d)$} was calculated using Oracle PGX software with a longest shortest distance of the largest network component\cite{RN134, RN143}.  The geodesic distance $d_{ij}$ between any two vertices $(i,j)$ is defined as the length of the shortest path between them. The network diameter $d$ is defined as $\textrm{max}~d_{ij}~\forall~(i,j)$.

\item \textit{Network degree assortivity $(r)$} is defined by the degree assortativity coefficient which has the form: 
\begin{displaymath}
r = \frac{\sum_{ij} (A_{ij} - k_i k_j/2m)k_i k_j}{2m - \sum_{ij} (k_i \delta_{ij} - k_i k_j /2m)k_i k_j}.
\label{eq:assortivity}
\end{displaymath}
Here $m$ is the total number of edges, ${\bf A}$ is the adjacency matrix encoding the connectivity structure, $k_i$ refers to the degree of vertex $i$, and $\delta_{ij}$ is the Kronecker delta function~\cite{RN140, RN143}.   Assortativity is a measure of whether like vertices connect to like vertices (in this case those with similar degree). This measure (which is formally equivalent to the Pearson correlation coefficient) lies in the range $-1 \leq r \leq 1$, with negative values associated with disassortative mixing (i.e high degree vertices more often connected to low degree vertices) and positive values with assortative mixing (i.e. similar degree vertices more often connected to each other).   In many networks (e.g. social networks), vertices tend to be connected to others with similar degree values\cite{RN151}.

\item \textit{Network reciprocity (\(\rho\))} is defined as the fraction of reciprocal edges over all edges in a directed graph, where $v_i \rightarrow v_j$ and $v_j \rightarrow v_i$ constitute a reciprocal pair\cite{RN135}.  In a directed graph, this provides a measure of how many bidirectional connections there are in a network.  A low reciprocity in a directed healthcare network may suggest that patients only flow in one direction between two healthcare organizations $v_i \rightarrow v_j$, without minimal reciprocal flow, such as in hospice care referrals for terminally ill patients. In an undirected network, reciprocity is trivially 1 for all pairs of vertices by definition.

\item \textit{Global clustering coefficient (C) or transitivity} We can quantify the level of transitivity in a network as follows. If $u$ knows $v$ and $v$ knows $w$, then we have a path $uvw$ of two edges in the network. If $u$ also knows $w$, we say that the path is closed--it forms a loop of length three, or a triangle, in the network. In the social network jargon, $u$, $v$, and $w$ are said to form a closed triad. We define the clustering coefficient to be the fraction of paths of length two in the network that are closed. That is, we count all paths of length two, and we count how many of them are closed, and we divide the second number by the first to get a clustering coefficient $C$ that lies in the range from zero to one~\cite{RN136, RN143}.   A high clustering coefficient in our networks can result when most providers are connected to other providers within the network, for example in a group practice that shares patients between providers.

\item \textit{Network density (D)} is calculated as $d m/{n(n-1)}$ where $n$ is the number of vertices, $m$ the number of edges, and $d = 1$ if the graph is directed or $d = 2$ if the graph is undirected\cite{RN137,RN138}.  Network density provides a measure of how tightly connected elements of the network graph are, and is essentially a ratio expressing the number of actual edges between vertices to the number of possible edges if the network were a complete graph (e.g. all vertices are connected to all other vertices).  This gives a measure of how interconnected the entire set of healthcare providers or organizations are.  At a city or regional level, network density may be quite high, but we might expect a low network density for the country as a whole since. Providers on either cost are not likely to share many patients and therefore not be connected by edges.

\item \textit{Largest component size (lco)} is the number of vertices in the largest connected graph component \cite{RN141, RN143}.  Some graphs may have several components (e.g. groups of edges) that are discontinuous, containing no common connecting edges.  This is a measure of network fragmentation, as networks in which the \textit{lco} is a small fraction of the total vertex count.  In other cases, the \textit{lco} is the dominant component containing the vast majority of vertices.

\item \textit{Betweenness centrality (\(C_{\beta})$} is calculated for an individual vertex $v_k$ and is the number of shortest paths between all pairs of vertices that go through $v_k\)\cite{RN142, RN143}.  For comparison between graphs, we also calculate (\(C^{\prime}_{\beta})$, which is the normalized betweenness centrality such that  $C^\prime_\beta = C_\beta \slash(N-1)(N-2)$ for directed networks, and  $C^\prime_\beta = 2C_\beta \slash(N-1)(N-2)$ for undirected networks.  A provider with a high (\(C_{\beta})$ value might be an oncologist, who receives referrals from many providers, but also refers patients to oncologic surgeons, radiation oncologists, hospice care, and many other types of providers.

\end{itemize}

\subsection*{High Performance and Parallel Computing Environment}

Analyses were run on BlueHive2, an IBM parallel cluster located at the Center for Integrated Research Computing of the University of Rochester.  We generally used two compute nodes, each with 2 Intel Xeon E5-2695 v2 processors with 12 cores and 64 and 512 GB of physical memory.  Network analysis was performed using Oracle Labs Parallel Graph Analytics (PGX) toolkit version 1.2.0 and Wolfram \textit{Mathematica} version 11.0 parallel computing and graph analysis functions.  \\

\section*{Results}
Our focus here is the comparison of topology and properties of the healthcare network graphs built using three algorithmic methods:  (1) a sliding temporal frame algorithm similar to that currently used to construct Medicare networks by the Center for Medicare Services\cite{RN118,RN980}, (2) a temporal binning method which captures all possible relationships within a given time span (e.g. creates a complete graph of all providers who saw the patient), and (3) a trace-route algorithm\cite{RN113,RN114} that builds networks based on sequential sequence of provider visits.  We have deliberately used networks generated from the Medicare Part B 2013 Outpatient Claims Limited Data Set, comprised of over 160 million claims, as opposed to a smaller data set.  Our motivation was not only to describe the differences in the topology of networks created by the three algorithms but also to investigate how these methods differ when used to address significant, real-world questions. For example: Which method will provide the most effective network representation for provider team identification or patient network flow? 

\subsection*{Comparison of graph metrics}

We first compared topological properties of network graphs constructed from the 2013 Medicare Part B Claims Data with the sliding frame, binning, and trace-route algorithms (Table~\ref{table2}). For this comparison, we used network graphs with $\tau=\)365 days. Medicare Part B Claims data files contain insurance claims for all outpatient Medicare encounters in the United States over the course of a year.  They do not contain charges for medications or hospitalizations.  Provider vertices are individual providers that provided and billed for care during the data set period of 2013.  Organization vertices represent provision of outpatient care by an organization.  In addition, providers are generally associated with or belong to organizations (e.g. a group practice), and each claim generally contains both a provider and their associated organization NPI number.  Because provider-provider networks (PPN) and organization-organization networks (OON) may have different network topologies and properties, and to separate the organization and provider dependencies of vertices, we constructed and analyzed separate networks for PPN and OON.

\begin{table}[!ht]
\begin{adjustwidth}{0.0 in}{0.0 in} 
\setlength{\tabcolsep}{8pt}
\caption{ 
\bf{Characteristics for patient co-care networks generated by different algorithms with $\tau=$365 days.}}
\begin{tabular}{l r r r c r r r}
\hline
\\ 
\vspace{2mm} {} & \multicolumn{3}{c} {Provider-Provider Networks} & \multicolumn{1}{c}{ } & \multicolumn{3}{c}{Organization-Organization Networks} \\
\cline{2-4} \cline{6-8} \\
\vspace{2mm} {Metric } & \multicolumn{1}{c} {Sliding} &\multicolumn{1}{c} {Binning\textsuperscript{\ddag}} &\multicolumn{1}{c} {Trace-route} &\multicolumn{1}{c} {}  & \multicolumn{1}{c} {Sliding} &\multicolumn{1}{c} {Binning\textsuperscript{\ddag}} &\multicolumn{1}{c} {Trace-route} \\
\hline  \\
Edges (E) & 89,377,290 & 65,287,590 & 40,077,297 & & 3,282,133 & 2,233,601 & 2,014,859 \\ 
Edge Type & Directed & Undirected & Directed & & Directed & Undirected & Directed \\ 
Vertices (V) & 811,784 & 811,784 & 814,917 & & 40,749 & 40,749 & 40,768 \\ \\
$ E_{loop} /E $ \textsuperscript{\dag}& -  & -  & 0.411  & & -  & -  & 0.122  \\ 
$ V_{loop} /V $ \textsuperscript{\dag} & -  & -  & 0.938  & & -  & -  & 0.943  \\  \\
$ d $ & 51 	& 29	& 89 	& 	& 6 & 13 	& 10\\ 
$ r $ & 0.05534 	& 0.06985 	& 0.02521 & 	&0.14768 	& 0.16542 	& 0.15915 \\
$ \rho $ & 0.56975 	& 1.0 	& 0.77929 	& 	& 0.86637 	& 1.0	& 0.97295 \\
$ C $ & 0.28097 	& 1.0	& 0.21598  	& 	& 0.53721 	& 1.0	& 0.57809\\ 
$ D $ & 0.00014 	& 0.00010 	& 0.00005  	& 	& 0.00198 	& 0.00135	& 0.00119\\ 
$ lco $ & 811,099 	& 811,099	& 810,952 	& 	& 40,749 	& 40,749 	& 40,749\\  \\
Max. V degree. & 19,320 	& 12,836 	& 10,857 & 	& 8,905 	& 5,248 	& 6,485 \\ 
Mean V deg. & 126.4 & 12.13 	& 67.34  	& 	& 7.982	& 2.057 	& 3.364\\
Max. E weight. & 75,985 	& 3,128 	& 22,166  	& 	& 376,808 	& 32,039 	& 472,774\\ 
Mean. E weight.  &  7.982	& 2.057 	& 3.364  	& 	& 126.4 & 12.13 	& 67.34  \\ \\ \hline
\end{tabular}
\begin{flushleft}
\vspace{2mm} $d$: network diameter, $r$: assortivity, $\rho$: reciprocity,  $C$: global clustering coefficient, $D$: network density, $lco$: number of nodes in the largest component.  \textsuperscript{\ddag}Metrics for undirected graph, \textsuperscript{\dag}Algorithm explicitly excludes self-loops.
\end{flushleft}
\label{table2}
\end{adjustwidth}
\end{table}

\begin{figure}[!ht]
\centering
\includegraphics[width=0.7\linewidth]{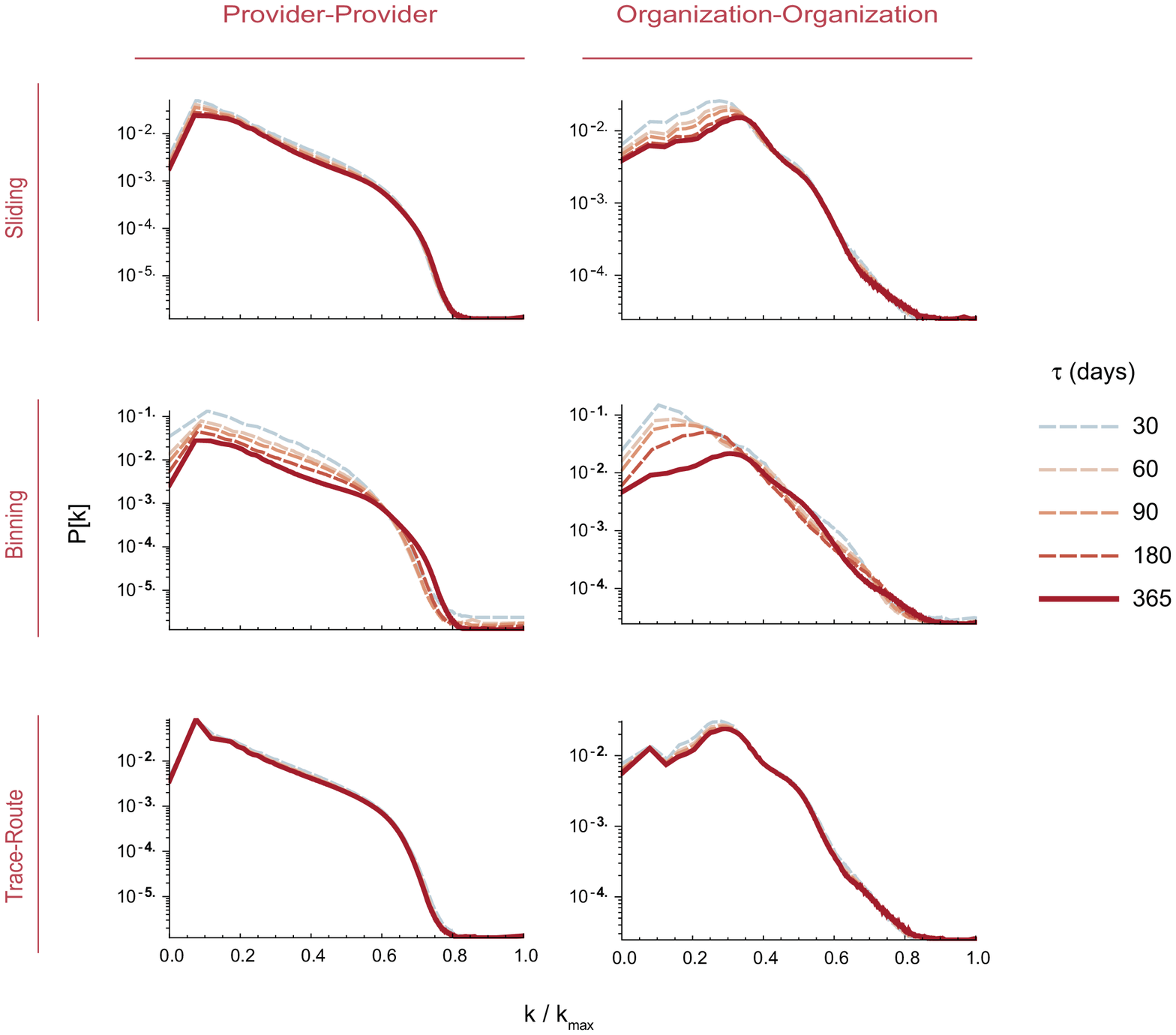}
\caption{{\bf Network stability with different sampling frames.} To assess network stability at various $\tau$, we plotted normalized vertex degree ($k/k_{max}(\tau)$) versus $P(k)$.  Very small variations in the plots at $\tau=$30, 60, 90, 180 and 365 days indicate that network properties as a function of $k/k_{max}(\tau)$ do not vary substantially.}
\label{PLfigure}
\end{figure}

All three algorithms yielded sparse networks, with the trace-route method having the lowest density values. All algorithms also selected similar numbers of vertices $V$, with the trace-route algorithm producing modestly more vertices due to inclusion of degenerate self-loop edges $(v_i \rightarrow v_i)$, representing sequential visits to the same provider.  In contrast, the binning and trace-route algorithms resulted in PPN with markedly fewer edges (73\% and 44\% less respectively) compared with the sliding method, along with a higher graph density and and maximum vertex degree. The large components $(lco$) were essentially of identical size across all three methods, and for both PPN and OON graphs. In order to check the variation of the degree distribution $P(k)$ with temporal frame $\tau$ we plot the rescaled degree $k/k_{max}(\tau)$ in function of $P(k)$ finding that for both the sliding frame and trace route algorithms, the vertex degree distribution properties are virtually identical for all $\tau$. We do note, however, some variation for the binning method at low $k/k_{max}(\tau)$ (Fig ~\ref{PLfigure}).  While these results give confidence that the algorithms capture virtually identical sets of providers or organizations, the large variation in the number of edges $E$ resulted in correspondingly large variations in network properties.

The binning algorithm generates non-directed graphs, and thus cannot be used to detect reciprocal events between providers with different edge weights (e.g. $v_k \rightarrow v_l$ coupled with $v_l \rightarrow v_k\)), or to infer directionality of provider-provider interactions.  However, the binning algorithm generates a complete provider graph for each patient, which makes it ideal for capturing complete ``teaming" or for identifying larger communities of providers.  In contrast, the sliding frame algorithm can also have multiple identical provider-provider or organization-organization edges (pairing weighted) for each patient, giving larger graph mean and maximum edge weights. This results from the sliding frame algorithm counting the same teaming interaction multiple times even if these are not sequential.    

The trace-route algorithm yielded the smallest networks in terms of edge counts, primarily because edges are counted only when the visits between providers were sequential in time.  The PPN created with the trace-route algorithm had a high fraction of edges that were self-loops (e.g. $v_i \rightarrow v_i\)).  Self-loops were present in 41.1\% of all edges and 94\% of all vertices in PPN and OON created with the trace-route algorithm.  This reflects the common pattern where a patient will see the same provider in succession multiple times.  Degenerate self-loops are not captured by the the binning or sliding frame algorithms.  This is a key issue when creating networks to model patient flow through healthcare systems.  If a large proportion of visits are sequential and to the same provider, algorithms that do not include degenerate self-loops cannot be used to accurately estimate network flow or capacity.  The sliding window algorithm, similar to that used by the Center for Medicare Services to generate publicly available Medicare networks, does not have this feature.

\begin{figure}[!ht]
\centering
\includegraphics[width=.8\linewidth]{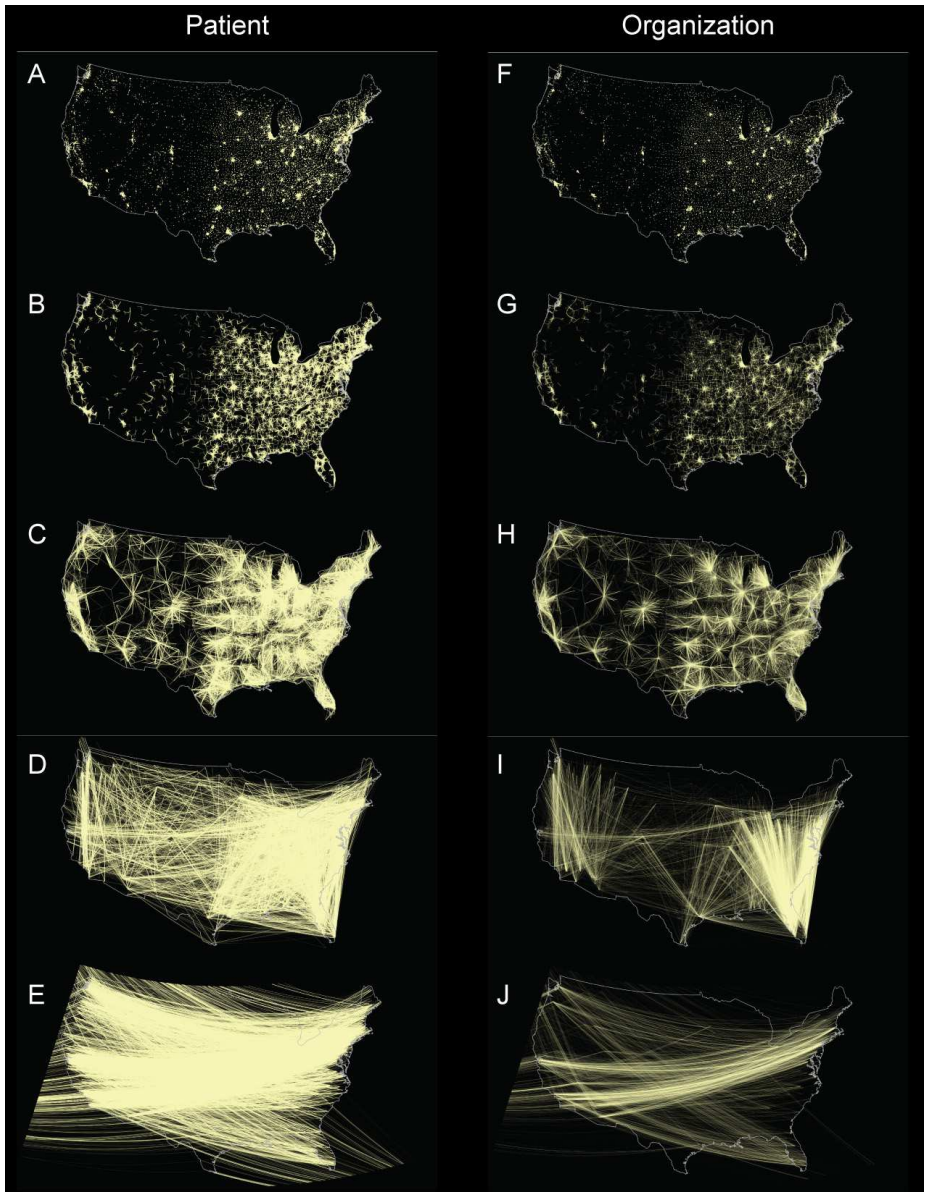}
\caption{{\bf Geographic visualization of healthcare networks.} Provider-provider (A-E) and Organization-Organization (F-J) healthcare networks for the Medicare Part B 2013 Limited data set created using the trace-route algorithm with a temporal frame $\tau=365$, and plotted using a geographic layout tied to provider location. The graphs are censored by removing weighted edges with a value of 1 (only a single shared patient) and excluding them from the visualization giving edge counts of 9,267,241 for PPN and 808,358 for the OON.  Each PPN image contains $\sim$ 1--9 million edges, and each OON image contains $\sim$ 0.2-2 million edges.  Images are binned by distance between providers:  $\leq 1$ mile (A,F), 20-40 miles (B,G), 100-200 miles (C,F), 800-1000 miles (D,G), 2000-4000 miles (E,H).}
\label{fig2}
\end{figure}

In order to uncover some of the spatial regularities associated with the constructed networks, we show a representative set in Fig~\ref{fig2}, plotted with a geospatial layout.  These networks were created with the trace-route algorithm, and each edge represents a sequential pair of visits between two provider.  This is contrast to networks built with the sliding frame or binning algorithms, where edges do not represent sequential visits (i.e. two providers may have an edge despite the patient never having seen them in immediate succession).  Given the rather large number of edges in the networks, for visualization purposes, we excluded edges with weights $\Omega_j$=1, which decreased the number of plotted edges for PPN by 76.9\% and for OON by 59.9\%.  To further enhance the resolution of the visualizations, edges were sorted in ascending order by the distance between two providers that constituted the vertices of an edge, and then plotted in 16 separate network subsets of approximately by the geospatial distance between vertices (i.e. providers or organizations).

The supplemental figures contain high-resolution geospatial network plots (using identical thresholding) for PPN and OON created by the binning (\nameref{s1_Fig}), trace-route (\nameref{s2_Fig}), and sliding frame (\nameref{s3_Fig}) algorithms, respectively. These plots contain 16 figures for each combination of PPN or OON with each algorithm.  Each of the 16 sub-plots contains a set of edges binned by the geographic-distance between provider or organization vertices in the edges (e.g. 2-4 miles, 4-10 miles, etc.), allowing a direct geospatial comparison of each method across plots.

There are several noteworthy features visible in the networks constructed by the trace-route method. The first is that the majority of edges appear to have very short distances ($\leq 10$ miles), suggesting that most Medicare patients have a set of providers in close proximity to each other.  This seems likely a result of the Medicare population mix, individuals over 65 years of age, on dialysis, or with disabilities from complex medical conditions \cite{RN144}, as well as a general preference not to travel long distances from home for medical care. Another striking feature is the density of healthcare providers and organizations in the Eastern and Midwestern states, which correlates well with population density.  In addition, note the spoke-and-hub appearance (Fig.~\ref{fig2}B,C,G,H) in both PPN and OON. This pattern appears to reflect travel between rural and urban areas.  Importantly, these edges do not reflect hospitalizations, which are not contained in Medicare Part B claims data.  Another notable feature is the presence of sequential visits between providers in the Northeast and the State of Florida (Fig.~\ref{fig2}D,F).  These likely represent the ``snowbirds", patients who spend winters in Florida.  The migratory nature of this group is reflected in the large north-to-south group of edges.  

Furthermore, one can also see key differences between the PPN and OON in Fig.~\ref{fig2}. Notably the differences in edge counts are apparent in the visible edge densities, around $\sim$40 million for the PPN and $\sim$2 million for the OON. High resolution images for both the PPN and OON constructed using all three algorithms, censored and uncensored, are  available in SI, Figs. S1--S3.Our preliminary investigations, thus indicate the valuable insights that one can glean through a geospatial representation of healthcare networks as it relates to patient journeys and flows. 

\subsection*{Censoring by edge weight markedly decreases network size}

We next examined the effect of censoring edges with low edge weights.  A key principle in public release of healthcare network data is to prevent identification of any individual patient, even within unipartite PPN or OON projections of bipartite networks where individual patients are not identified as vertices.  An individual patient might be identified by a combination of unique providers they see where the edge weights between the majority of those providers is 1, and each provider can be identified by a geographic area.  Such convergence to unicity (i.e. the ability to identify an individual from a unique combination of attributes) only requires a small number of attributes in very large data sets\cite{RN112}. Unfortunately, this may lead to fragmented networks, with many small sub-networks unconnected to the largest connected component.  Medicare censors edges in publicly released provider teaming data sets by excluding those with weights \textless 11 ("Presumed shared relationships based on claims for fewer than eleven distinct beneficiaries will be excluded from the report.") \cite{RN118}.

\vspace{1mm}
\begin{table}[!ht]
\setlength{\tabcolsep}{2pt}
\caption{
\vspace{1mm}
\bf{Effect of edge weight censoring\textsuperscript{\ddag}}}
\begin{tabular}{l r r@{\hskip .4cm}  r@{\hskip .4cm} r r@{\hskip .4cm}  r@{\hskip .4cm}  r@{\hskip .4cm} r r r r r}
\hline
\\ 
\vspace{0.5mm} {}  & \multicolumn{1}{p{0.4cm}}{ } & \multicolumn{3}{c} {\bf{Uncensored}} &
\multicolumn{1}{p{0.4cm}}{ } & \multicolumn{3}{c}{\bf{Censored}} &\multicolumn{1}{p{0.4cm}}{ } & \multicolumn{1}{c}{} &\multicolumn{1}{p{0.4cm}}{ } & \multicolumn{1}{c}{}\\
\cline{3-5} \cline{7-9} \\
\vspace{2mm} & \multicolumn{1}{c}{ } & \multicolumn{1}{c} {$lco$} & \multicolumn{1}{c} {$V_{N}$} &\multicolumn{1}{c}{$\displaystyle \frac{V_{N}}{V}$} &  \multicolumn{1}{c}{ } & \multicolumn{1}{c} {$lco$} & \multicolumn{1}{c} {$V_{N}$} &\multicolumn{1}{c} {$\displaystyle \frac{V_{N}}{V}$} & \multicolumn{1}{c}{} & \multicolumn{1}{c}{\bf{$ \displaystyle \frac{V_{Censored}}{V_{Uncensor}}$}} & \multicolumn{1}{c}{} & \multicolumn{1}{c}{\bf{$ \displaystyle \frac{E_{Censored}}{E_{Uncensor}}$}}\\
\hline  \\
\bf{Provider} \\
Sliding &  & 810,099 & 685  & $<$0.001 &  & 191,414 & 30,545 & 0.1376 & &0.273 & & 0.017\\ 
Binning &  & 810,099 & 685  & $<$0.001 &  & 229,054 & 24,358 & 0.0961 & &0.312 & & 0.045  \\ 
Trace-route &  & 810,952 & 3,963  & 0.005 &  & 87,533 & 228,444 & 0.7230 & &0.388 & & 0.018 \\
\\
\bf{Organization} \\
Sliding &  & 40,749 & 0  & - &  & 35,733 & 44 & 0.001 & &0.878 & & 0.102 \\ 
Binning &  & 40,749 & 0  & - &  & 36,338 & 30 & $<$0.001 & & 0.892 & & 0.092 \\ 
Trace-route &  & 40,749 & 19  & $<$0.001 &  & 36,680 & 1,966 & 0.054 & &0.899 & & 0.140 \\
\\
\hline
\end{tabular}
\begin{flushleft}
\vspace{2mm} {$lco$}: vertices in largest component, {$V_{N}$}: vertices not connected to largest component,  {$V$}: total number of vertices,  {$V_{Censored}$}: total number of vertices in networks with censoring of edges with weights $\leq\)11, {$E$}: total number of edges in uncensored networks;  {$E_{Censored}$}: total number of edges in networks with censoring of edges with weights $\Omega_{v_j \rightarrow v_k} \leq\)11, {$V$}: total number of vertices in uncensored networks; \textsuperscript{\ddag}Metrics for undirected networks with $\tau=\)365 days.
\end{flushleft}
\label{table4}
\end{table}

To examine the effect such censoring has on the resulting network properties, we compared uncensored and censored provider and organization networks created by each algorithm.  We hypothesized that such censoring would lead to network fragmentation.  We found that censoring, compared to uncensored networks, resulted in a striking reduction in both nodes and edges (Table~\ref{table4}), as well as network density.  This was most evident with respect to edges, where censoring for $\Omega_{v_j \rightarrow v_k} \leq \)11 resulted in removal of more than 95\% of edges for PPN, and more than 85\% of edges for OON.  There is currently no standard for labeling these edges "noise" with the true "signal" being the edges with $\Omega_{v_j \rightarrow v_k} > \)11.  For example, a provider may have a moderate number of Medicare patients, say 300, who see 30 different specialists (e.g. surgeons, oncologists, endocrinologists, etc.).  If each subspecialist sees less than 11 patients, the edges will not appear in the PPN. This substantial reduction in network structure strongly suggests that censored networks will not provide a full picture of healthcare network topology.

\subsection*{Comparison of power-law characteristics of healthcare networks}

\begin{figure}[!ht]
\centering
\includegraphics[width=0.9\linewidth]{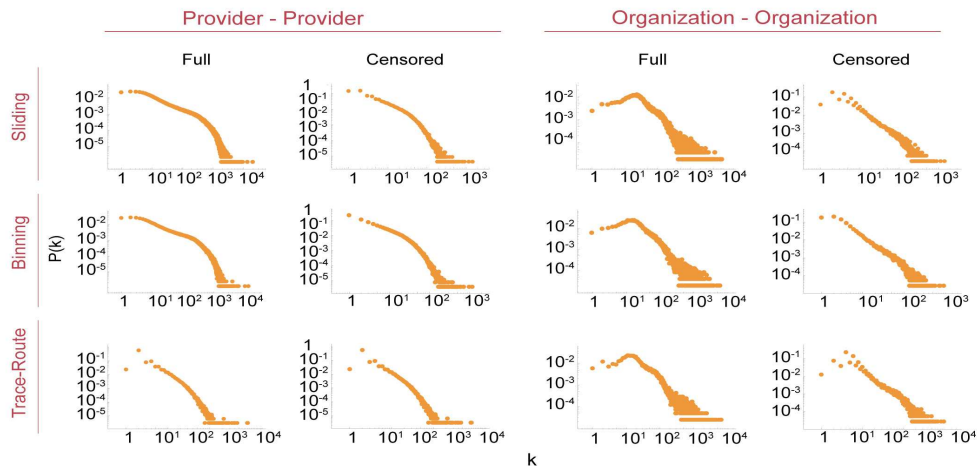} \\
\caption{{\bf Network vertex degree distribution by algorithm for $\tau=365$}}  
\label{fig4}
\end{figure}

We next tested networks generated by these methods to determine whether they were scale-free that adhered to a vertex-degree power law distribution.  Many large and sparse networks are scale-free\cite{RN89, RN125, RN111, RN146}, with power-law characteristics indicating a small number of central hubs with many edges, and a small-world topology\cite{RN111}. Networks that can be described by power law distributions have distinct properties that have implications for network formation and evolution\cite{RN89,RN143}. In the case of healthcare networks, for example, power law behavior may suggest how networks grow. For example a doctor in a new medical practice is  likely to refer patients to other highly established providers with many connections, a phenomenon known as preferrential attachment in graph theory \cite{RN89}.  In contrast, Medicare healthcare organizations are state-based, are more likely to be linked to other networks within or immediately adjacent to their state, and have a degree distribution pattern that obeys a discontinous power law \cite{RN124}. For the PPN and OON built in this manuscript, vertex degrees $k$ and their frequencies $P(k)$, are shown in Figure~\ref{fig4}.  Neither PPN or OON appeared to obey a strict power law distribution (e.g. $f(x) \propto x^{-\alpha}$). Interestingly, uncensored OON had a $P(k)$ distribution similar to that found in relatively high density networks of internet discussion groups \cite{RN126}.  Censoring by edge weight $\omega<\)11, however, decreased network density $D$, and altered the $P(k)$ distributions in all networks. 

 We then used the method of Clauset et al. \cite{RN111} to test PPN and OON network degree distributions for goodness of fit with a power law distribution, and to compare the fit with other discrete distributions (power law with exponential cutoff, exponential, log normal, Weibull, and Yule).  If the power law had $p>0.1$ then we accepted the null hypothesis that the data followed a power law distribution.  To determine if the plausibility was significant we used the likelihood ratio (LR) to compare the fit with one of the other discrete heavy tailed distributions listed above\cite{RN111}. If the LR values were negative with $p<0.05$, we concluded that the network followed the alternative distribution being tested rather than the the power law distribution. The distribution with the most negative LR was selected as the best fit.

\begin{figure}[!ht]
\begin{adjustwidth}{0 in}{0 in} 
\centering
\includegraphics[width=1.\linewidth]{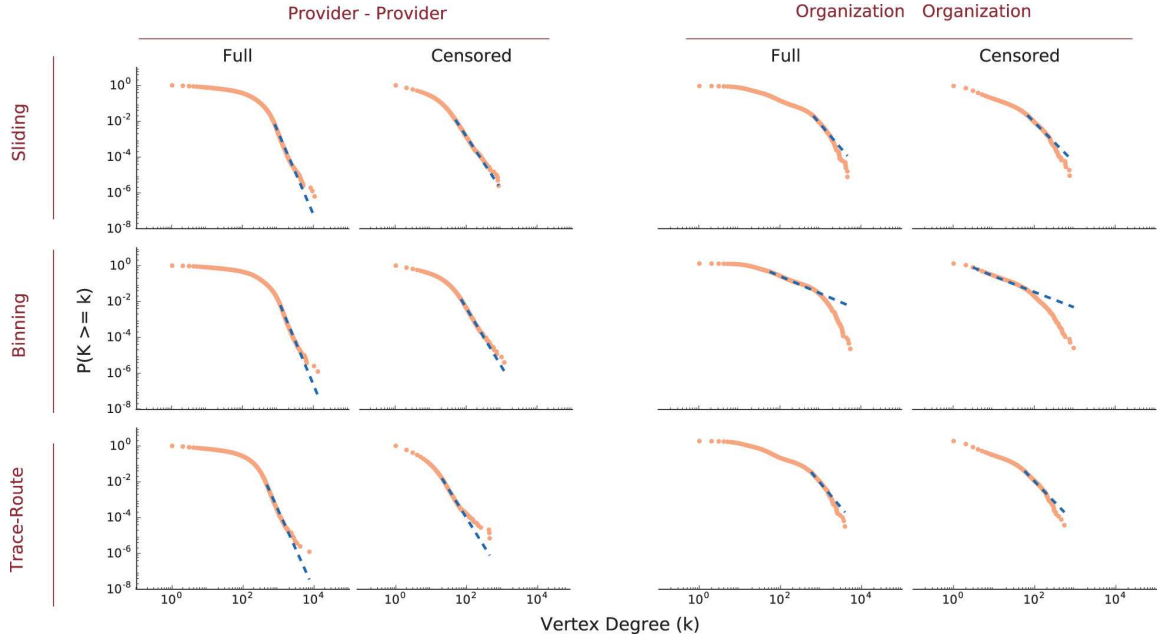} \\
\vspace{3mm}
\caption{{\bf Healthcare networks power law with cutoff properties}  We analyzed PPN and OON for adherence to power a law distribution starting from a minimum vertex degree, $(x_{min})$, using the method of Clauset, et al \cite{RN111} with $\tau=\)365 days. The orange points are the CDF of the vertex degree, and the blue dashed line is the power law fit of the CDF given $x_{min}$ and $\alpha$.}
\label{fig5}
\end{adjustwidth}
\end{figure}

In our analysis (Figure~\ref{fig5}), all but one of the full and censored PPN adhered to the power law distribution with thresholding, that is beyond a value of $x_{min}$, with statistical significance Table~\nameref{S2 Table}. Goodness of fit testing for the Poisson distribution yielded likelihood ratio results that were approximately the order of magnitude of 5 (i.e. $10^5$), and hence insignificant. However, none of the other heavy tailed distributions fit the full PPN created with the sliding frame algorithm, and all had an LR$>$0.

In contrast to PPN, the best goodness of fit on the OON was the power law with exponential cutoff (PLEC; Table~\nameref{S2 Table}).  This distribution is often seen in network analysis of human mobility \cite{RN124}, reflecting an opportunity cost for traveling larger distances.  Similarly, the spatial distributions of for-profit and public facilities obey a PLEC distribution, which is consistent with previously described models where there is a higher financial cost for locating facilities in areas of sparse population \cite{RN125}.  In the case of OON, the better fit to a PLEC distribution suggests that shared numbers of patients between healthcare organizations decay proportional to the distances between organization service areas.  This seems logical, given that most healthcare providing organizations (e.g. hospitals, clinics) are regionally based.

\subsection*{Vertex centrality distribution varies by network construction algorithm}

\begin{figure}[!ht]
\centering
\includegraphics[width=1.\linewidth]{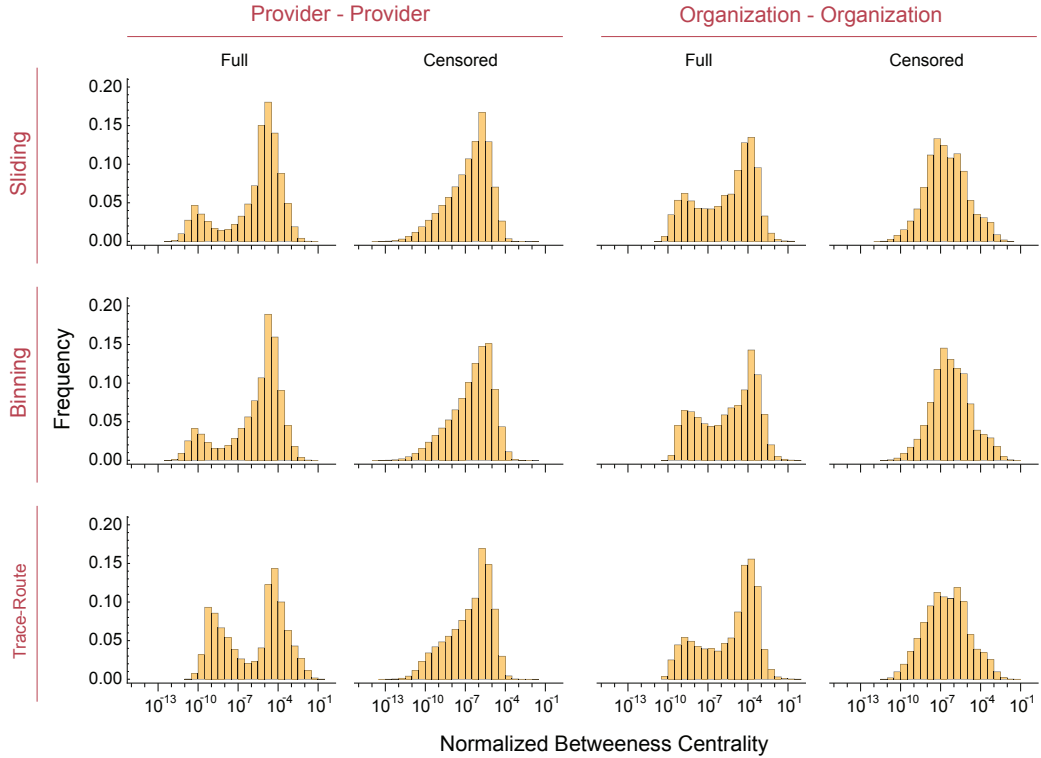} \\
\caption{{\bf betweenness centrality $C^\prime_\beta$ of healthcare networks by algorithm for $\tau\)=365}  betweenness centrality was calculated for all networks using the Oracle PGX algorithm.  Results are displayed with algorithmic binning of $C^\prime_\beta = C^\beta \slash(N-1)(N-2)$ for directed graphs produced by the sliding frame and trace-route algorithms, and  $C^\prime_\beta = 2C^\beta \slash(N-1)(N-2)$ for undirected networks produced by the binning algorithm.  All plots are scaled in the y-axis to frequency, allowing direct comparison of centralities.  Note that edge-weight censoring (excluding edges with $\Omega_{v_j\rightarrow v_k} \leq$ 11) markedly changes the centrality distribution of all networks.}
\label{fig4A}
\end{figure}

We next analyzed differences in the centrality and connectivity of individual vertices (providers or organizations) between the networks generated by the three algorithms.  Centrality metrics may be used to rank organizations by the proportion of shared patients with many other organizations, and can also  be used to analyze healthcare service provision disparities or revenue potential. (Figure~\ref{fig4A})  Normalizing betweenness centrality (\(C^\prime_\beta$ - see Methods), and plotting the frequency rather than absolute distribution, allows direct comparison of all the networks despite their differing size and scales.  

The $C^\prime_\beta$ distributions of networks produced by the different algorithms are quite similar.  The major difference is between full and censored networks.  The addition of nodes that are only connected by edges with $\Omega_{v_j\rightarrow v_k} \leq$ 11 introduces a bimodal distribution of $C^\prime_\beta$, reflecting the low centrality of the previously censored nodes.  This is consistent with the hypothesis that most of these providers or organizations are on the periphery of the network backbone, and are unlikely to create new connections between other providers or organizations that already have high $C^\prime_\beta$ or $k_i$ values.  Another possibility is that the this phenomenon reflects variation in the proportions of total patients with Medicare insurance seen by providers.  Some providers may see a large percentage of Medicare patients (e.g. nephrologists and geriatric medicine practitioners), while others may see only a small number of Medicare patients but a much higher proportion of patients with private insurance, leading to a bimodal distribution of $C^\prime_\beta$.

\subsection*{Network variation by temporal sampling frame interval}

\begin{figure}[!ht]
\centering
\includegraphics[width=1 \linewidth]{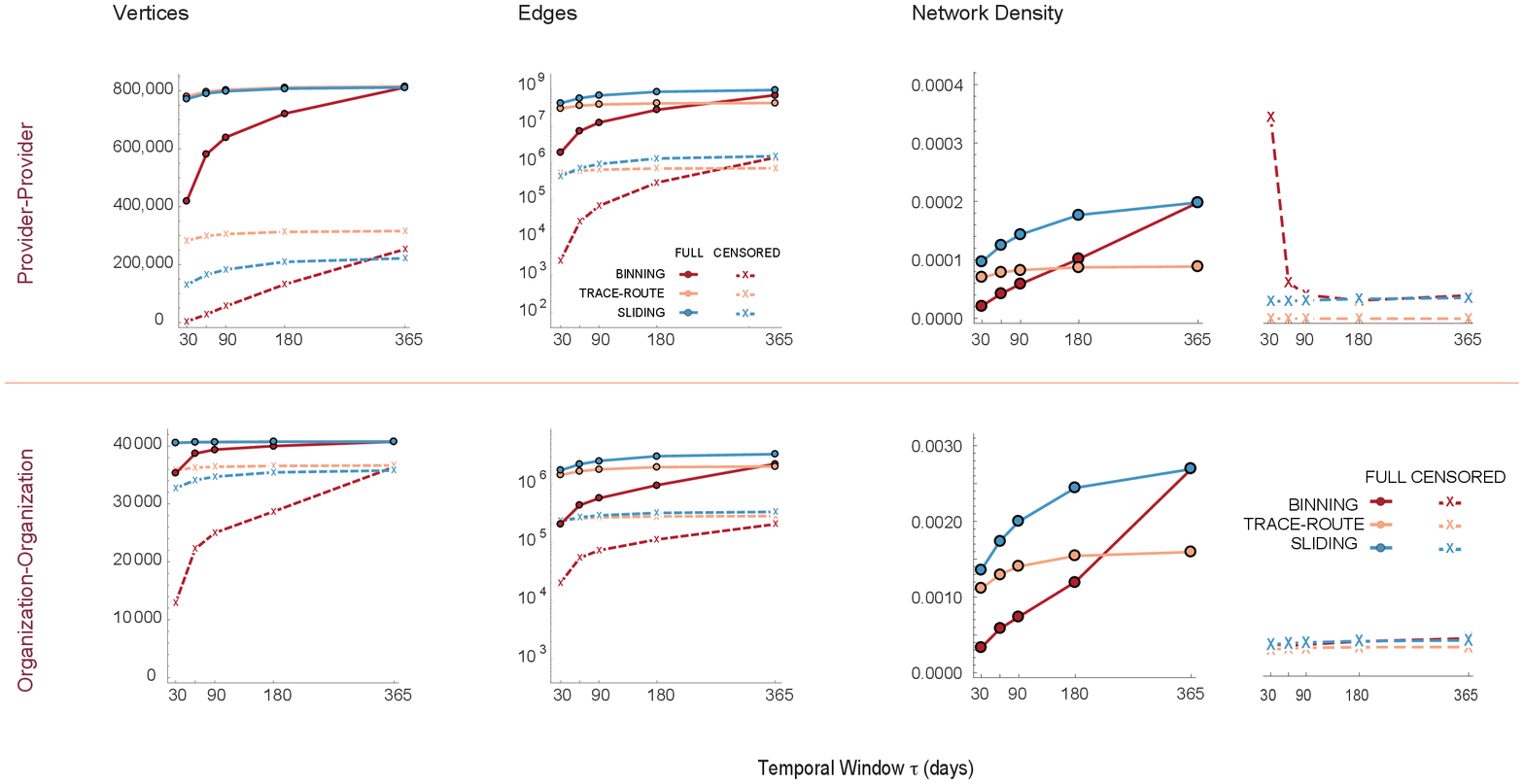} \\
\caption{{\bf Network vertex counts, edge counts and density as a function of the sampling frame interval $\bf\tau$}  Vertex counts, edge counts and network density plotted for provider and organization networks for the binning (red), trace-route (orange) and sliding frame (blue) algorithms for $\tau$ = 30, 60, 90, 180, and 365 days.  Solid lines represent networks where vertices were included if the minimum edge weight $>= 1$, while dashed lines represent censoring where only edges with a minimum edge weight $>= 11$ are included.  The latter is the standard for aggregate data release by the Center for Medicare Services so that individual patients cannot be identified by a unique combination of providers sharing only a single patient.
}
\label{fig67}
\end{figure}

Healthcare networks are dynamic; the  shared number or volume of patients between providers (e.g. edge weights) changes over time based on frequency of patient visits and, less frequently, as new providers are added to or leave the network.   Of particular interest has been the number of patients shared between two providers during a specific time period.  This measure might reflect the efficiency of patient flow through a healthcare system.  For example, if the number of edges reflecting shared provider visits within 30 days is small, it might suggest that patients are unable to get urgent consultations in a timely manner.  Alternatively, examining networks built from visits during a time frame $\tau$ can help determine if longer $\tau$ provide a fuller picture of the network (Figure~\ref{fig67}). We found that over $\geq$ 98\% of vertices and edges are captured with  $\tau  \geq$ 90 days for both PPN and OON by the trace-route and sliding frame algorithms (see Table S1).  Considerable variation, however, occurs with the binning method, and the number of vertices or edges included converges on that of the binning and trace-route methods only as $\tau$ approaches 365 days.  These findings suggest that the major topology of claims-derived healthcare networks can be captured with $\tau\)=180 days. 

\subsection*{Variations in network community identification}

One use of PPN is to identify highly collaborative teams or communities of providers.  Such communities can arise from shared patient patterns (e.g. referrals) or membership in financial organizations (e.g. Accountable Care Organizations, practice networks, or group practices).  Provider teams can be identified by network community identification algorithms \cite{RN146}, as well as hierarchical or agglomerative clustering methods\cite{RN9,RN11}.  The composition and number of groups will vary by method, and is a function of vertex connections via edges and edge weights.  Thus, it is highly likely that networks built using different algorithms, when analyzed by the same community identification method, will yield different groups of providers.  To test this hypothesis, we examined community assignments resulting from networks generated using the same data by each algorithm. 

\begin{figure}[!ht]
\centering
\includegraphics[width=1\linewidth]{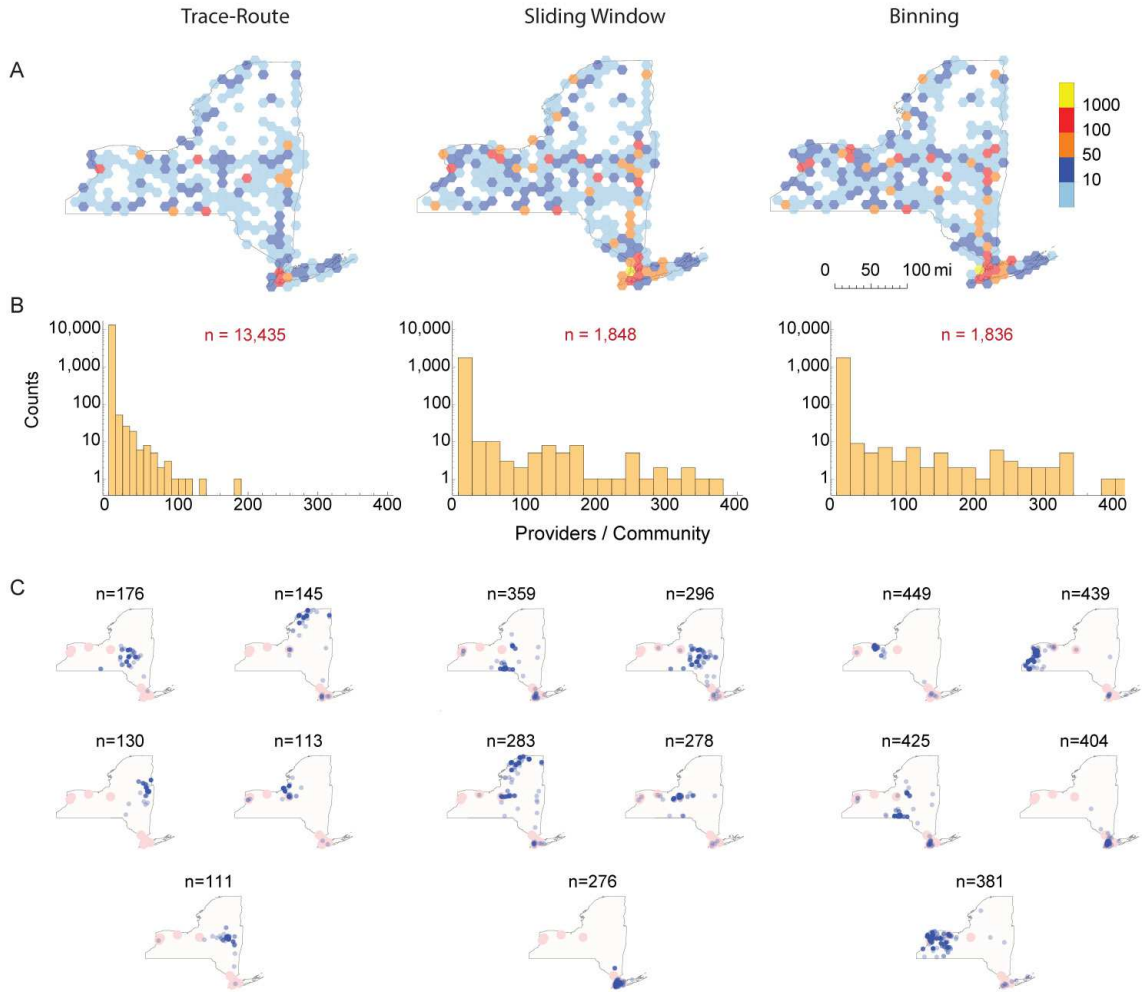} \\
\caption{{\bf Variation in provider community identification}.  We analyzed undirected Provider-Provider networks constructed with the trace-route, sliding frame and binning algorithms for $\tau$ = 365, and censored for edge weights $\le 11$.  Provider-Provider community teams identified for providers within NY State from each network using the Girvan-Newman modularity community finding algorithm \cite{RN146} implemented in \textit{Mathematica}.  Each provider was assigned to only one community. (A) Provider densities.  Hexagonal bins show the counts of providers that were members of any community within each geographic region color coded by range.  Note the different geographic density patterns for each method.  (B) Histogram of number of providers per community. Note the large number of communities ($n$) in each histogram, with the majority having only 2 providers.  Community sizes, compositions and number differed between all 3 methods.  (C) Shows the five largest communities identified in each network. 
}
\label{figCom1}
\end{figure}

Figure \ref{figCom1} illustrates how networks built from the same data set using the trace-route, sliding frame, and binning methods yield different provider communities.  A community is defined as a set of vertices (e.g. providers) who have a larger number of connections (e.g. shared patients) with each other than vertices outside the community. For this analysis, we started with PPN for $\tau$ = 365, and censored for edge weights $\le 11$.  For simplicity of comparison and computational efficiency, we  selected only edges where both providers were located in NY State.  Communities were identified by the Girvan-Newman modularity community finding algorithm \cite{RN146}.   We found marked variation in the number, size, and composition of the resulting communities.  Similar results were found when this procedure was applied to other states.  Not only is the number and geographic distribution of providers belonging to communities different (\ref{figCom1}A and B), but the community size and geographic distribution ranked by number of providers also differs substantially.  The community partitioning of the trace-route networks yielded a large number of small communities (97\% with $n \le$ 6) compared to the sliding frame and binning method networks (46$\%$ and 44\% with $n \le$ 6 respectively).  The geographic location of the communities, when ranked from largest to smallest, also differed substantially (\ref{figCom1}C).  This analysis highlights the significant differences seen in identifying provider communities when networks are built with different algorithms. For example, the trace route method has a lower representation of providers in New York City in the largest 5 communities than the binning or sliding frame algorithms.

\section*{Discussion}

Healthcare networks are commonly constructed from insurance claims data to study several activities:  referrals, teaming, and communication.  These are used to identify network topology, patterns of provider association, and to test whether these correlate with healthcare outcomes or as part of comparative effectiveness research.  The results of such analyses are increasingly used to shape healthcare services delivery and policy, with potential to impact almost a million providers and over 40 million Medicare Part B beneficiaries. However, little work has been published on the effect of network algorithm selection on network topology and analysis.  Our results demonstrate that different algorithms will yield different results, and that that algorithm selection should carefully consider several factors:  the "meaning" of edges and edge weights, the temporal sequence and frame used for analyzing claims, and whether to censor PPN data to prevent individual patient identification when provider locations and details are known.

A critical step in healthcare network analysis is defining the "meaning" of an edge, and selecting a congruent algorithm for network building.  The trace-route algorithm has the most obvious interpretation, where the resulting network edges represent the total sequential visits of all shared patients between two providers.  Alternatively, edges in networks constructed with the binning method represent shared patients between providers.  Less clear is the meaning of edges created by the sliding frame algorithm, which is currently used by Medicare to produce the annual Medicare Physician Referral Data Set \cite{RN980, RN113,RN118}.  Weighted edges created by the sliding frame method have been thought to represent provider-specialist referrals, the specific act of sending a patient from provider $v_k$ to provider  $v_l$.  This interpretation is problematic for several reasons.  For example, in the United States, many health plans permit patients to select a specialist without requiring a referral by a primary care provider.  In addition, not every provider-provider pairing constitutes a referral;  a patient alternating visits between their internist and their oncologist may or may not constitute multiple referrals to the same provider.  Our analysis suggests that the sliding frame algorithm does not capture such nuances, and inflates the number of "referrals" between providers. 

Another related and important facet of meaning is the mistaken inference of causality.  Referrals are a real event in medical care, but these cannot be identified solely by the temporal order of visits or providers.  For example, both the trace-route and the sliding window methods use temporal ordering of claims.   However, this is simple temporal sequence and does not imply causality.  For example, a patient may have been to their oncologist, then 5 days later to the Emergency Room, next had a yearly physical with their Family Practitioner, and then a follow up to the ER visit some days later.  Here we are just capturing that sequence.  The oncologist may not have referred the patient to the ER five days later (e.g. after an automobile accident unrelated to their breast cancer 5 year follow up), and the ER visit did not "cause" the physical.  Similarly, we should be careful inferring that teams identified in network analysis imply conscious teaming of specific providers.  The choice of providers may be highly determined by external factors such as geography (e.g. there is only one abortion provider within 300 miles) or payment constraints (e.g. the patient's insurance will only reimburse for visits with specialists within the insurance network).  Thus, network analysts should be extremely cautious in using network structure alone to infer voluntary teaming for laudable goals (e.g. providing the best care for prostate cancer) or more sinister intent (e.g. insurance fraud or inappropriate narcotics prescriptions). 

Our findings indicate that the trace-route and binning algorithms may provide more consistent mappings between meaning and network structure.  The trace-route algorithm uses sequential visits between providers to build edges, and the resulting network can be viewed as a map of directed patient flow between provider visits through the Medicare system.  It is similar to algorithms used to map the Internet, and shares features such as temporal ranking of paths, decomposition of routes into edges, and network edge weights that represent aggregate flow between vertices\cite{RN113,RN114}.  It also captures self-loops, the common occurrence of sequential visits to the same provider, an event otherwise excluded by the currently used sliding frame algorithm.  Trace-route networks are commonly used to model packet flow through computer networks\cite{RN116}, vehicle flow through highways\cite{RN115}, and the flow of goods through supply chains\cite{RN149}. Patients are not packages, but such methods do provide a way to study patient flow and efficiency in healthcare networks.  In contrast, binning networks capture provider teaming and identify linked communities even if patients do not sequentially visit all providers within their team.  Our results indicate that provider community identification may be best performed on networks constructed with the binning method to identify tightly linked provider teams and communities.

Another key finding is difference between the vertex degree distribution of OON and PPN. That is, the OON tend to follow a power law with exponential cutoff, while the PPN tend to follow a power law. This is likely due to geographic constraints; Medicare is administered at the state level, and providers are geographically based.  This suggests that at some point there is limited value for organizations to extend their interactions (e.g. share patients) over long distances. Practically, this limits the vertex degree of healthcare organizations, and their regional associations.  In contrast, providers in PPN appear to obey a preferential attachment rule; they tend to add connections exponentially such that new providers tend to link with established and well connected providers.  These differences suggest underlying structural differences in the social, structural and economic motivations of providers and organizations, and that further work to better define these could be done comparing different healthcare systems (e.g. single versus multi-payer systems).

Given the above considerations, our findings support the use of specific algorithms for network construction depending on the intended analysis:  using the trace route algorithm when analyzing network patient flow, and the binning algorithm for identifying provider or organization communities or teams.  Analysis of uncensored networks is preferred, as censoring dramatically reduces network edges and alters network topology.  In addition, mixing provider and organization vertices is can be problematic, depending on the type of analysis undertaken, given the dependencies between organizations and providers (e.g. provider connections are highly dependent on organization membership), and the differences in network topology between PPN and OON. Bipartite network analyses of how providers are linked to organizations would be appropriate for a mixed provider-organization graph, but provider-provider teaming, topology or flow analyses are probably best undertaken with separate PPN and OON networks.  Together, these findings imply that analyses based on the current publicly available Medicare networks (as of September 2016)\cite{RN67, RN71}, created using an implementation of the sliding frame algorithm\cite{RN118,RN980}, may have significant flaws:  there is no clear interpretation of the meaning of the edges, networks are censored for low frequency edge weights, and they mix provider and organization vertices and edges which may be problematic for some analyses.  Studies of the US Medicare healthcare network should strongly consider building networks from primary Medicare claims data using verifiable open source algorithms, and analyzing uncensored networks.  Finally, our work suggests several future directions of inquiry, including comparisons of networks between states with different Medicare structures, composition of provider teams, studies to identify provider migration to different geographic locations, and studies of patient flow through healthcare systems.
 
\section*{Conclusion}

The topology of healthcare networks constructed from claims data varies as a function of the algorithms used to construct them.  Consequently, the analytic results obtained will vary accordingly, including network density, edge weights, vertex centrality measures, and community identification.  Good practice for healthcare network analysis should include building networks from primary claims data, analyzing uncensored networks, and using explicitly defined algorithms.   The choice of algorithm is highly significant and should be matched to the questions being addressed.  From this study, we conclude that the trace-route algorithm is most suited to analyses of patient flow through the network, and the binning algorithm to studies of provider teaming and community identification.  

\section*{Supporting Information}


\paragraph*{S1 Fig.}
\label{s1_Fig}
{\bf Healthcare network plots created with the binning algorithm.} Provider-provider and organization-organization network plots for $\tau=365$ as discussed in the Results section.  Networks are plotted using geospatial coordinates accurate to within 0.8 miles in the continental United States.  Each sub-plot represents a range of distances for edges between providers or organizations to allow comparisons across methods.  Figures are also available online:  PPN: \href{https://figshare.com/s/1054f48521ee7f75b715}{doi: 10.6084/m9.figshare.3827220}, OON: \href{https://figshare.com/s/8e4c310b512d5c285c03}{doi: 10.6084/m9.figshare.3827217}.

\paragraph*{S2 Fig.}
\label{s2_Fig}
{\bf Healthcare network plots created with the trace-route algorithm.} Provider-provider and organization-organization network plots for $\tau=365$ as discussed in the Results section. Figures are also available online:  PPN:  \href{https://figshare.com/s/ef453524484445b9f3e3}{doi: 10.6084/m9.figshare.3827532}, OON:  \href{https://figshare.com/s/eb15be0bed3fd0b71f82}{doi: 10.6084/m9.figshare.3827520}.

\paragraph*{S3 Fig.}
\label{s3_Fig}
{\bf Healthcare network plots created with the sliding window algorithm.} Provider-provider and organization-organization network plots for $\tau=365$ as discussed in the Results section. Figures are also available online:  PPN \href{https://figshare.com/s/638bd98a64c59c620978}{doi: 10.6084/m9.figshare.3827505}, OON:  \href{https://figshare.com/s/7c68005ef9d19a2ab6b7}{doi: 10.6084/m9.figshare.3827361}.

\paragraph*{S4 Fig.}
\label{s3_Fig}
{\bf Healthcare community identification for NY state.}  Provider-provider communities with n>5 providers were identified in networks built for New York State providers only.  These plots show all of the provider locations for each identified community. Major cities are identified in red. Figures are also available online: \href{https://figshare.com/s/638bd98a64c59c620978}{doi: 10.6084/m9.figshare.3827505}.

\paragraph*{S1 File.}
\label{S1_File}
{\bf Network construction algorithms}  This file contains the sliding, trace-route and binning PERL algorithms for network construction. (doi: 10.6084/m9.figshare.3837717)

\paragraph*{S2 File.}
\label{S3_File}
{\bf Censored networks}  Censored network files are available from figshare.com at \href{https://figshare.com/s/915649f63d8fe8b08c5e}{doi: 10.6084/m9.figshare.3833943}

\paragraph*{S1 Table.}
\label{S1Table}
{\bf Table of graph metrics as a function of measurement frame $\tau$.} This table, in TSV format, contains all of the the metrics for 200 networks built using the binning, trace-route, and frame algorithms for providers or organizations, with or without censoring, directed or undirected edges, and edge weights reflecting shared patients, or total numbers of visits for the shared patients.

\paragraph*{S2 Table.}
\label{S2 Table}
{\bf Power law best-fit results for patient co-care networks with $\tau =$ 365 days}  This table contains the statistical testing results for network fitting.

\section*{Acknowledgments}

The authors would like to thank Fred Trotter, Orna Intrator, Ann Dozier and Katia Noyes for spirited discussions on healthcare networks that greatly enhanced this manuscript.  This work was supported in part by grants: U54 TR001625 from the National Center for Advancing Translational Sciences (NCATS), and a grant from the Philip Templeton Foundation.

\nolinenumbers


\newpage
\noindent
{\large \bf{Supplemental Table 1}}

\begin{table}[!ht]
\begin{adjustwidth}{0 in}{0 in} 
\setlength{\tabcolsep}{1pt}

\caption*{
\vspace{1mm}
\noindent
\bf{Power law best-fit results for patient co-care networks with \(\tau=\)365 days\textsuperscript{\dag}}}
\begin{tabular}{l r r r r r r r r r r r r r r r r r}
\hline
\\ 
\vspace{0.5mm} {}  & \multicolumn{1}{p{0.2cm}}{ } & \multicolumn{1}{p{0.2cm}}{ } & \multicolumn{2}{{p{2cm}}} {PLEC} &
\multicolumn{1}{p{0.2cm}}{ } & \multicolumn{2}{p{2cm}}{Exponential} & \multicolumn{1}{p{0.2cm}}{ } & \multicolumn{2}{p{2cm}}{Log normal} & \multicolumn{1}{p{0.2cm}}{ } & \multicolumn{2}{p{2cm}}{Weibull} & \multicolumn{1}{p{0.2cm}}{ } & \multicolumn{2}{{p{2cm}}}{Yule} \\
\cline{4-5} \cline{7-8} \cline{10-11} \cline{13-14} \cline{16-17} \\
\vspace{2mm} & \multicolumn{1}{c}{PL-\(p\)} & \multicolumn{1}{c}{ } & \multicolumn{1}{c} {LR} &\multicolumn{1}{c} {\(p\)} &  \multicolumn{1}{c}{ } & \multicolumn{1}{c} {LR} &\multicolumn{1}{c} {\(p\)} &  \multicolumn{1}{c}{ } & \multicolumn{1}{c} {LR} &\multicolumn{1}{c} {\(p\)} &  \multicolumn{1}{c}{ } & \multicolumn{1}{c} {LR} &\multicolumn{1}{c} {\(p\)} &  \multicolumn{1}{c}{ } & \multicolumn{1}{c} {LR} &\multicolumn{1}{c} {\(p\)}   \\
\hline  \\
\bf{Provider-Provider} \\
Sliding (Full)  & 0.011 &  & 0 & 1 & & 400 & 1 & & 0.232 & 0.984 & & 25.16 & 1 & & 0.088 & 0.931\\ 
\vspace{3mm} Sliding (Censored)  & \textbf{0.560} & & 0 & 1 & & 365 & 1 & & 0.018 & 0.554 & & 11.62 & 0.980 & & -0.083 & 0.425 \\
Binning (Full) & \textbf{0.231} & & 0 & 1 & & 189 & 1 & & 0.141 & 0.971  & & 152 & 1 & & 0.044 & 0.935 \\ 
\vspace{3 mm} Binning (Censored) & \textbf{0.928} & & 0 & 1 & & 227 & 1 & & 0.029 & 0.677 & & 8.751 & 0.971 & & 0.060 & 0.588 \\ 
trace-route (Full) & \textbf{0.521} d& & 0.036 & 1 & & 240 & 1 & & 0.152 & 0.971 & & 18.83 & 0.998 & & 0.113 & 0.946 
\\
trace-route (Censored) & \textbf{0.301} &  & 0 & 1 & & 260 & 1 & & 0.076 & 0.973 & & 10.56 & 0.992 & & 1.260 & 0.92  \\\\
\bf{Org-Org} \\
Sliding (Full)  & 0 &  & \textbf{-26.66} & \hspace{1mm} $<$0.001 & & 0.767 & 0.522 & & -24.55 & \hspace{1mm} $<$0.001 & & -24.89 & $<$0.001 & & -0.110 & $<$0.001 \\ 
\vspace{3 mm}Sliding (Censored)  & 0 & & \textbf{-15.64} & $<$0.001 & & 56.49 & 1 & & -12.85 & 0.001 & & -13.83 & 0.001 & & -0.658 & $<$0.001 \\
Binning (Full) & 0 & & \textbf{-330} & $<$0.001 & & 3500 & 1 & & -192 & $<$0.001  & & -219 & $<$0.001 & & -3.965 & $<$0.001 \\ 
\vspace{3 mm} Binning (Censored) & 0 & & \textbf{-399} & $<$0.001 & & -230 & \hspace{1mm} $<$0.001 & & 8731 & 1 & & -210 & \hspace{1mm} $<$0.001 & & -63.35 & $<$0.001 \\
trace-route (Full) & 0.043 & & \textbf{-9.10} & $<$0.001 & & 6.474 & 0.756 & & -7.911 & 0.007  & & -7.488 & 0.059 & & -0.046 & $<$0.001 \\ 
trace-route (Censored) & 0.252 & &  \textbf{-7.18} & $<$0.001 & & 23.38 & 0.982 & & -5.568 & 0.016 & & -6.109 & 0.015 & & -0.339 & \hspace{1mm} $<$0.001 \\\\
\hline
\end{tabular}
\begin{flushleft}
\vspace{2mm} PL-\(p\): Power Law p-value Calculated by fitting 1,000 random power law curves, where the null hypothesis is that the data adheres to power law, LR: Likelihood Ratio, \(p\): p-value (Whether distribution is a better than a power law), PLEC:  Power law with exponential cutoff. Statistically significant values are in bold face type. \\ \textsuperscript{\dag}Algorithm explicitly excludes self-loops. \textsuperscript{\ddag}Metrics for undirected graph. 
\end{flushleft}
\label{stable1}
\end{adjustwidth} %
\end{table}

\end{document}